\documentstyle[aas2pp4]{article}

\lefthead{Fukugita and Hata}
\righthead{Metal Abundance in the Solar Interior}
 
\begin{document}

\title{Metal Abundance in the Solar Interior}
\author{M.\ Fukugita and N.\ Hata}
\affil{Institute for Advanced Study, Princeton, NJ 08540, U. S. A.}

\begin{abstract}

It is shown that the metal abundance in the solar interior is constrained
from the current solar neutrino experiment, even if one allows neutrino
conversion in the sun due to neutrino oscillation. The result shows that
the metal abundance in the interior should be within the range 0.4--1.4
times that in the surface, supporting the idea that the
sun formed by a contraction of a gas cloud with an almost homogeneous 
composition.

\end{abstract}

\section{Introduction}

Whether the metal abundance measured in the stellar surface represents
correctly the value in the stellar core is a nontrivial
question. In computations of the evolution of stars it is 
generally assumed that the initial metal abundance of the core is 
identical to the present day abundance of the surface, for which 
spectroscopic information is available
for heavy elements. 
This agrees with the simple picture that stars formed by a contraction
of the molecular cloud of the uniform composition.

In principle, however, the metallicities of the core and the 
surface can be different, unless
the whole stars are fully convective like in low mass ($M<0.4M_\odot$) 
main sequence stars (e.g., Iben 1967). An extreme example is Jupiter,
where the core consists mostly of silicates, while the surface is
dominated by hydrogen and helium with small admixture of heavier
elements. It may well be that the origin of Jupiter like stars
is completely different from that of ordinary stars (e.g., Podolak,
Hubbard \& Pollack 1993): nevertheless it 
would not be absurd to imagine that rocks are taken into the core
at the time of gas contraction, and the core has metallicity 
higher than in the surface since the birth of the star, for which
the radiative transport dominates.  There is also an effect
that heavy metals tend to sink by diffusion towards the core, 
while the envelope becomes hydrogen rich. A calculation shows that
such an effect do exist in the sun as evidenced from helioseismology, 
although the amount is small
(Bahcall et al. 1997).

The main reason that one assumes the homogeneous composition 
throughout the star is basically due to 
lack of information of the metal abundance in the stellar
interiors. In this article
we consider the problem that the sun might offer a possibility to 
study the metal abundance in the deep interior or in the core
through the solar neutrino observations that probe the core
region of the sun. This problem, however, is not very obvious
due to the famous solar neutrino problem that the observed neutrino
flux is by a factor 2--4 times smaller than is predicted with the standard 
model of the sun (Bahcall 1989). The most elegant and widely 
accepted solution to
this problem is that electron neutrinos emitted in the nuclear reactions
of the sun are
converted by neutrino oscillation into muon or tau neutrinos 
that are sterile in the nuclear
detector or have a much smaller cross section for scattering off
electrons, as advocated by Mikheyev and Smirnov (1986a; refereed to as 
the MSW effect).

This solution at first glance appears to make the things so flexible 
that almost any amount of the neutrino flux before the oscillation effect
is experimentally allowed, if one tunes the neutrino mass and mixing 
parameters in some appropriate way. 
Indeed, forgetting about all knowledge on the nuclear
reaction cross sections, even the case that almost 100 \% of the solar
energy is generated by the CNO cycle is not excluded solely by
the solar neutrino experiment (Bahcall, Fukugita \& Krastev 1996). 

This, however, is too extreme. If we adopt the knowledge of nuclear
reactions within the range allowed by the current experiment, the
freedom is not that large. For instance, the amount of $^8$B must be
in the range between 1/3--2 times the value the
standard solar model predicts,
and the core temperature can be determined to within 5\% of the standard
solar model value (Hata \& Langacker 1997; hereafter HL97) in order
to satisfy the current solar neutrino experiments, whatsoever the 
neutrino parameters one takes. In  this
paper we study to what extent the current solar neutrino experiments
constrain the metal abundance in the sun allowing for MSW neutrino
oscillation (or neutrino conversion).

The increase of the metal abundance obviously promotes the 
CNO cycle. It also increases opacity that modifies the core temperature.
Therefore, it increases the relative importance of the neutrino flux from
the CNO cycle significantly. This increment of the CNO neutrino flux 
must be cancelled by increasing importance of the
suppression factor coming from the neutrino oscillation to keep
the consistency with the solar neutrino experiments.  In so far
as this works, a larger metal abundance in the solar interior
is allowed. If the metal abundance is increased more, however, we 
no longer have 
solutions that satisfy all the solar neutrino experiments in a consistent
way.  This is the logic that we are going to explore in this study.
 
We remark that metal abundance may not be so tightly constrained from 
helioseismology information.
Helioseismology is sensitive to the sound velocity, the change of which
reads approximately $\Delta c_s/c_s\sim
1/2(\Delta T/T-\Delta \mu/\mu)$ with $T$ and
$\mu$ temperature and mean molecular weight,
$\mu^{-1}=2X+0.75Y+0.58Z(^{12}{\rm C})+...$.
A cancellation takes place between temperature and mean molecular
weight.  
Although the present best solar model is known to give $c_s$ as accurate
as 0.2\%, and the 50\% change in $Z$ might be detected if it changes
with satisfying $\Delta Z=-\Delta X$, the presence of $Y$ complicates
the situation. Hence, it is not obvious how strong constraint can be derived on
the metallicity independently of $X$ and $Y$.  

\section{Calculation}

We take the standard solar model of Bahcall and collaborators as our
fiducial (Bahcall \& Ulrich 1988; Bahcall \& Pinsonneault 1992; 1995,
hereafter BP95), 
and consider a small departure from their best model. The energy of the sun is 
generated from the $pp$ chain (98\%) and the CNO cycle (2\%).
Neutrinos are produced in $pp\rightarrow de\nu $,
$pep\rightarrow d\nu$, $e+^7$Be$\rightarrow^7$Li$+\nu$ 
and $^8$B$\rightarrow^8$Be$+e+\nu$ from the $pp$ chain, and in beta
decay of $^{13}$N, $^{15}$O and $^{17}$F in the CNO cycle. 
The experimental information
comes from the long-running Homestake experiment with $^{37}$Cl, which is
sensitive to both high energy neutrinos ($^8$B neutrinos) and 
intermediate energy neutrinos ($^7$Be, $pep$ and CNO neutrinos)
(Cleveland et al. 1997), a
water Cerenkov experiment at Kamiokande and Superkamiokande 
measuring only for high energy neutrinos (Fukuda et al. 1996; Totsuka et al.
1997), and gallium experiment, Gallex (Hampel et al. 1996) and
Sage (Abdurashitov et al. 1996) that are very sensitive 
to low energy neutrinos ($pp$ neutrinos). The problem is that the 
detection rate is smaller
than predicted, by factors, $3.7\pm0.6$, $2.6\pm0.5$ and $2.0\pm0.2$, 
respectively. Furthermore, this specific energy dependent
suppression pattern makes the explanation of the problem by modifying
the solar model highly unlikely, leaving the neutrino oscillation
explanation as the most attractive possibility (Bahcall \& Bethe 1990;
Fukugita \& Yanagida 1991).

Indeed, this energy dependent suppression is very naturally understood by 
considering the conversion of electron neutrinos into other types of 
neutrinos inside the sun by neutrino oscillation (Mikheyev \& Smirnov
1986a). 
For the neutrino flux given by the standard solar model, the current
solar neutrino experiments allow 
two small parameter regions that are located in the two parameter space, 
neutrino mass-square difference
$\Delta m^2=m_{\nu_e}^2-m_{\nu_i}^2$ ($i=\mu$ or $\tau$) 
and intrinsic mixing angle
($\theta$) between the two relevant neutrinos: 
one is called the small angle solution, in which intermediate energy neutrinos
are suppressed, and the other the large angle solution, for which the 
suppression of neutrino fluxes is almost energy independent.  The
most up to date calculations are found in Bahcall \& Krastev (1996)
and in HL97.

Additional information can be obtained from an upper limit on the
possible flux variation between the day and night time
(day-night effect) (Mikheyev \& Smirnov 1986b; see Bahcall \& Krastev 1997
and references therein). For 
some specific neutrino parameter range the converted 
muon neutrinos are changed 
back to electron neutrinos during the propagation through the Earth, 
causing an increase of the neutrino capture rate in night in the Kamiokande 
and Superkamiokande detectors.  The absence of this effect down to 2\%
level (Fukuda et al. 1996) excludes a substantial size of parameter 
regions of our interest.

We repeat the neutrino propagation calculation allowing for a variation
in the metal abundance of the sun.
We use the scaling law of Bahcall \& Ulrich (1988),
which has given the explicit metallicity dependence for each component of the 
neutrino flux. The range of the model explicitly studied covers about 
$\pm$50\% around the value of metallicity determined for the solar 
surface. The dependence 
outside this range is a simple extrapolation with power law. 
Although this calculation is rather old, 
we expect the gross metallicity dependence does not differ 
from what one could obtain from the more modern BP95 calculation. 
We impose a luminosity
constraint 
so that the luminosity that would 
change due to a change in the metal abundance
is renormalized to the today's luminosity of the sun. Namely, we study
the model at fixed luminosity. 

The table of Bahcall \& Ulrich shows that the most sensitive to metal
abundance are indeed the CNO neutrino fluxes: the power $\gamma$
of the flux $\phi\propto(Z/X)^\gamma$ is 1.86 for the $^{13}$N neutrino,
2.03 for the $^{15}$O neutrino and 2.09 for the $^{17}$F neutrino. This high 
power is caused by a multiplicative effect due to the increase of the 
abundance of catalysing $^{12}$C and the increase of opacity that makes
the core temperature higher. In spite of its sharp temperature dependence,
the effect on the $^8$B neutrino flux is smaller ($\gamma=1.27$) 
than for the CNO neutrinos.

As for the fiducial flux, we use the value of the BP95 calculation 
with metal diffusion effect taken into account:
$\phi(^8{\rm B})=(6.6\pm1.1)\times 10^6$ cm$^{-2}$s$^{-1}$,
$9.3\pm1.3$ SNU for captures with $^{37}$Cl and $137\pm8$ SNU
for captures with $^{71}$Ga.
We take 2.55$\pm0.14\pm0.14$ SNU for the Homestake experiment
(Cleveland et al. 1997),
($2.80\pm0.19\pm0.33)\times 10^6$cm$^{-2}$s$^{-1}$ for Kamiokande 
(Fukuda et al. 1996) and
($2.51{+0.14 \atop -0.13}\pm0.18)\times 10^6$cm$^{-2}$s$^{-1}$ for 
Superkamiokande [combined $(2.586\pm0.195) \times 10^6$cm$^{-2}$s$^{-1}$]
(Totsuka 1997),
and $69.5\pm6.7$ SNU for combined Sage and Gallex experiments.
The absence of the day-night effect is also imposed on our data analysis.  
In our actual calculation
we use the night-time flux data divided into 5 bins according to the cosine of 
angle from the sun (Fukuda et al. 1996).

The data are then fitted with the three free parameters, $\Delta m^2$,
$\sin^22\theta$ and $Z/X$, and calculate a likelihood function taking
account of both experimental errors of neutrino reaction rates and
those arising from uncertainties of solar models as given by BP95,
in the same way as done in HL97. 

The resulting probability distribution is displayed in Fig. 1 taking $Z$
as a parameter. The range allowed at 95\% confidence level (CL) is
$$0.4<Z/Z_{\rm surface}<1.4 \eqno{(1)}$$
where $Z_{\rm surface}=0.0175$ (Grevesse \& Noels 1993) using
the standard solar model value $X=0.71$. 
The region outside this range is excluded even if we assume the flux
reduction due to neutrino oscillation. Namely,
the metal abundance in the solar interior cannot be much different 
from that in the surface. The allowed range is, of course, much 
larger than the change of the metal abundance induced by the diffusion effect,
which is about 15\% in $Z$ (BP95).  We remark that the range given
in eq. (1) is the range where explicit solar model studies
are made by Bahcall and Ulrich (1988) and the behaviour regarding the variation
of $Z$ is well studied; so hindsight we need not to use power law
of the $Z$ dependence out to the range where its behaviour is not well
established.

\begin{figure}[t]
\epsscale{0.90}
\plotone{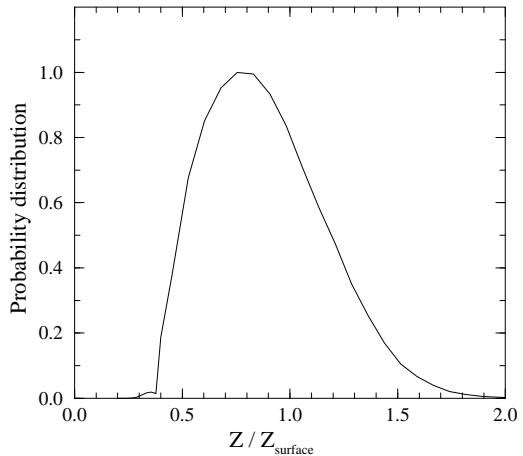}
\caption[Fig1.ps]{
Probability distribution as a function of metal abundance 
of the interior of the sun. Likelihood is normalized to the maximum height  
being unity (so that 95\% CL is at about 0.18).
\label{Fig1}}
\end{figure}

We also show a figure (Fig. 2) of allowed region for the two other parameters,
$\Delta m^2$ and $\sin^22\theta$, overlaid on the corresponding
figure with the standard case (i.e., $Z=Z_{\rm surface}$ allowing for
errors for the value at the surface 6.1\%). The contour is a parameter range
corresponding to 95\% CL. Compared with the contour of the standard
model case, the allowed region is elongated horizontally for
the small angle solution, or vertically
for the large angle solution. Most of the elongated parts 
(a part in the left hand side of the small angle solution, and
that in the upper part of the large angle solution)
correspond to the case with metallicity lower than the default value.
For larger metallicity, the change required for the neutrino parameters
is quite small to accommodate increased neutrino fluxes.
Nevertheless, there is a sharp cut off against the increase of metallicity,
beyond which appropriate neutrino parameters do not exist to make the
flux consistent with the {\it three} solar neutrino experiments. In our
calculation the constraint from the absence of the day night effect 
serves to squeeze the high metallicity end in the large angle solution,
but it plays little role for the small angle solution.

\begin{figure}[t]
\epsscale{0.82}
\plotone{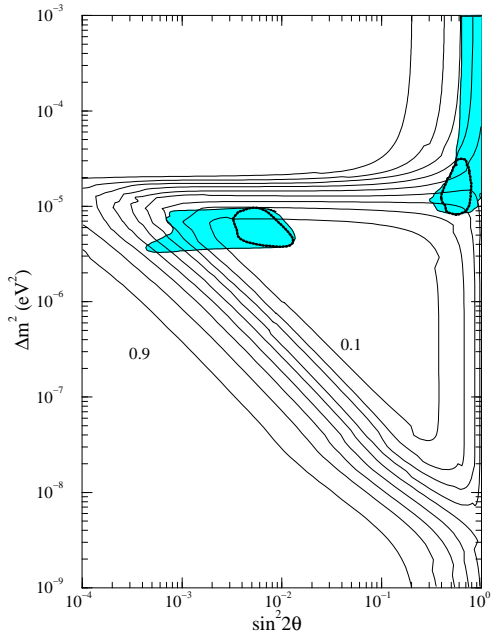}
\caption[Fig2.ps]{
Allowed neutrino parameters (95 \% CL) when 
metallicity is treated as a free parameter (shaded regions).  
The thick solid contours 
are the region allowed for the standard solar model ($Z=Z_{\rm surface}$) 
at 95 \% CL. The contours indicate the suppression factor for the
$^{13}$N neutrino flux in steps of 0.1 unit, as discussed in the
discussion section below.
\label{Fig2}}
\end{figure}

\section{Discussion}

We have shown that the solar neutrino experiments put a strong constraint
on the metal abundance in the interior of the sun, even if we allow the
neutrino oscillation due to the matter effect. Namely, the possibility of
neutrino oscillation does not lend us much freedom to increase metallicity
inside the sun. The metallicity in the interior of the sun 
should not be larger than the surface value 
by more than 40\%,
or 0.15 dex in [Fe/H] for the initial value.  
This is a good news to the people working for stellar evolution calculations,
since we expect that the sun is not a special case, but the same
probably applies to more general cases, justifying the standard 
assumption that stars formed by a contraction of
a homogeneous gas sphere (Hayashi 1966).

On the other hand, the errors in the current neutrino experiments and 
uncertainties in nuclear reaction cross sections still allow the 
possibility that the metal abundance in the solar interior is slightly 
larger (smaller) than in the surface. Accepting this uncertainty, 
energy generation from the CNO cycle may be as uncertain as
0.4\% to 4\% of total
energy generation, which is compared to 1.8\% for the standard value
(see Fig. 3). The $^{13}$N or $^{15}$O neutrino fluxes may vary
from 1$\times 10^8$ to 12$\times 10^8$cm$^{-2}$s$^{-1}$ within this
range.

\begin{figure}[t]
\epsscale{0.90}
\plotone{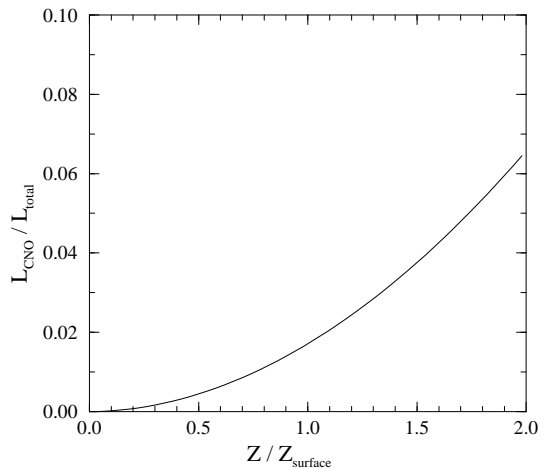}
\caption[Fig3.ps]{
Fraction of the solar energy borne by the CNO chain as
a function of the metal abundance in the solar interior.
\label{Fig3}}
\end{figure}

One might think that the $^{13}$N or $^{15}$O neutrino flux 
(the end point energies 1.199 or 1.732 MeV, respectively)
itself can give us
a useful indicator for the metallicity in the solar interior due to
its sensitivity  to the carbon abundance. One may prepare a detector
with a detection threshold set just above the energy of $^7$Be 
neutrinos (0.862 MeV).
Once a new detector, which can measure $^7$Be neutrinos, e.g., with
a liquid scintillator measuring for $\nu e\rightarrow \nu e$, is constructed
[e.g., Borexino (Arpesella et al. 1992)], 
this is not difficult, since the CNO neutrino flux is 100 times 
higher than $^8$B neutrino flux in the standard solar model.  
The oscillation effect, however, makes
the situation somewhat subtle. For the allowed regions of neutrino parameters,
we expect the MSW suppression factor for the $^{13}$N 
neutrino flux to be 0.1--0.5 (0.4--0.7)
for the small (large) angle solution  for a 
given metallicity. Unless this large uncertainty arising from
neutrino oscillation is reduced, the
$^{13}$N flux does not give us useful information on the metallicity.

\begin{figure}[t]
\epsscale{0.90}
\plotone{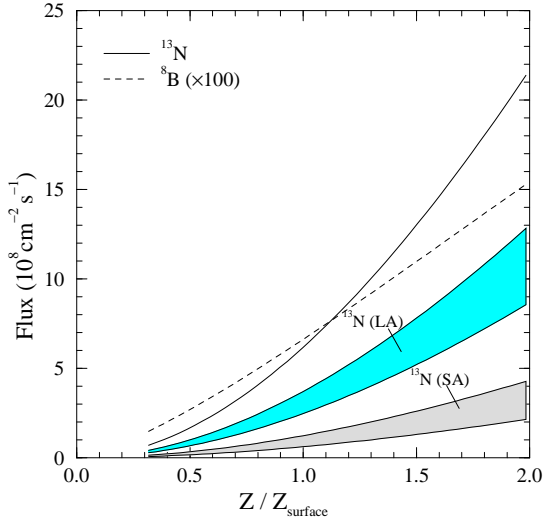}
\caption[Fig4.ps]{
$^{13}$N and $^8$B neutrino fluxes as a function of
the metal abundance in the solar interior. Solid and dashed curves 
indicate those fluxes without modifications due to neutrino conversion
(the flux of $^8$B neutrinos is multiplied by a factor of 100). The shaded
regions are those expected for $^{13}$N flux after neutrino conversion 
for the small angle (SA) and large angle (LA) solution.  We assume that
the error of gallium capture rate was reduced to $\pm2$ SNU and
the $^8$B neutrino neutral current reaction was measured 
with an error of 10\%.
\label{Fig4}}
\end{figure}

We find that the most effective way to reduce this uncertainty is to
measure gallium capture rate as precise as possible. If the error attached
to gallium capture rate is reduced, the region shrinks in the vertical
(horizontal) direction for the small (large) angle solution, i.e., the
error in the suppression factor is reduced.  
A $^8$B neutrino flux measurement via the neutral current interaction
planned at Sudbury, when combined with a 
better gallium experiment, serves to further reduce the error in the
suppression factor for a given metallicity (although this experiment alone is
not very effective for our purpose). 
For instance, if one achieves an error of the gallium experiment
as small as $\pm2$ SNU around the current best value and
measure the $^8$B neutrino neutral current reaction
to a 10\% accuracy, we would obtain 
$0.85<Z/Z_{\rm surface}<1.2$ at 95\% CL, if the small angle solution is
right, or $0.9<Z/Z_{\rm surface}<1.1$ if the large angle solution is correct
(see Fig.~4).

We have argued that solar neutrino experiments have given unique
information for the metals in the solar interior, which is not accessible
by other means. Assuming that the sun is not special, this removes 
our worry that metal abundance in the stellar interior might be different 
from that in the surface. For example, this can be an issue in deriving
metallicity dependence of the Cepheid period-luminosity relation, for
which the metal effect on the colour, which is affected by surface metal
abundance, cancels to a large degree against that on the luminosity,
which is affected by metal abundance in the core of the sun (Stothers
1988). If
metallicity would be different in these two places, the large cancellation
no longer takes place, resulting in a much larger metallicity dependence
in the Cepheid period-luminosity relation.  The present analysis, if
our result applies to star formation from molecular clouds in general,
implies that the metallicity difference, at least, is not a 
likely possibility to account for a
large metallicity dependence of the Cepheid period
luminosity relation suggested recently (Gould 1994; Sasselov et al. 1997; 
Sekiguchi \& Fukugita 1997).  
We have also discussed that one
could reduce the error in $Z/Z_{\rm surface}$ to a level of 10--20\%
with further improvement in some specific solar neutrino experiments.
At this level, the effect of metal diffusion could be seen with the
solar neutrino experiments, opening a possibility to carry out a cross check
with the result from helioseismology.

\acknowledgements

We would like to thank John Bahcall, Plamen Krastev and Bohdan Paczy\'nski
for very stimulating discussions and John Bahcall for valuable comments
on the preliminary draft. 
M.F. wishes to acknowledge support from
the Fuji Xerox Corporation. N.H. is supported by the National Science
Foundation Contract No NSF PHY-9513835.


\begin{thebibliography}{}

			

\bibitem[Abdurashitov et al. 1996]{Abdurashitov-etal-1996}

Abdurashitov, J. N. et al. 1996, Phys Rev Lett 77, 4708
					

\bibitem[Arpesella et al. 1992]{Arpesella-etal-1992}

Arpesella, C. et al. 1992, BOREXINO proposal, Vols.\ 1 and 2,
ed.\ by G.\ Bellini et al.\ (Milan: University of Milan) 
					

\bibitem[Bahcall 1989]{Bahcall-1989}

Bahcall, J. N. 1989, Neutrino Astrophysics (Cambridge: Cambridge
                University Press)


\bibitem[Bahcall \& Bethe 1990]{Bahcall-Bethe-1990}

Bahcall, J. N. \& Bethe, H. A. 1990, Phys Rev Lett 65, 2233


\bibitem[Bahcall, Fukugita \& Krastev 1996]{Bahcall-Fukugita-Krastev-1996}

Bahcall, J. N., Fukugita, M. \& Krastev, P. 1996, Phys Lett B374, 1


\bibitem[Bahcall \& Krastev 1996]{Bahcall-Krastev-1996}

Bahcall, J. N. \& Krastev, P. I. 1996, Phys Rev D53, 4211


\bibitem[Bahcall \& Krastev 1997]{Bahcall-Krastev-1997}

Bahcall, J. N. \& Krastev, P. 1997, Institute for Advanced Study 
                preprint IASSNS-AST 97/31


\bibitem[Bahcall \& Pinsonneault 1992]{Bahcall-Pinsonneault-1992}

Bahcall, J. N. \& Pinsonneault, M. H. 1992, Rev Mod Phys 64, 885


\bibitem[Bahcall \& Pinsonneault 1995]{Bahcall-Pinsonneault-1995}

Bahcall, J. N. \& Pinsonneault, M. H. 1995, Rev Mod Phys 67, 781


\bibitem[Bahcall et al. 1997]{Bahcall-etal-1997}

Bahcall, J. N., Pinsonneault, M. H., Basu, S. \& Christensen-Dalsgaard,
                1997, Phys Rev Lett 78, 171


\bibitem[Bahcall \& Urlich 1988]{Bahcall-Urlich-1988}

Bahcall, J. N. \& Ulrich, R. K. 1988, Rev Mod Phys 60, 297


\bibitem[Cleveland et al. 1996]{Cleveland-etal-1996}

Cleveland, B. T. et al. 1996, preprint


\bibitem[Fukuda et al. 1996]{Fukuda-etal-1996}

Fukuda, Y. et al. 1996, Phys Rev Lett 77, 1683


\bibitem[Fukuda \& Yanagida 1991]{Fukuda-Yanagida-1991}

Fukugita, M. \& Yanagida, T. 1991, Mod Phys Lett A6, 645


\bibitem[Grevesse \& Noels 1993]{Grevesse-Noels-1993}

Grevesse, N \& Noels, A. 1993, in Origin and Evolution of the Elements,
         ed. by  N. Prantzos et al. (Cambridge, Cambridge University Press),
         p.15 


\bibitem[Gould 1994]{Gould-1994}

Gould, A. 1994, ApJ 426, 542


\bibitem[Hampel et al. 1996]{Hampel-etal-1996}

Hampel, W. et al. 1996, Phys Lett, B388, 384


\bibitem[Hata \& Langacker 1997]{Hata-Langacker-1997}

Hata, N. \& Langacker, P. 1997, Institute for Advanced Study preprint
             IASSNS-AST 97/29



\bibitem[Hayashi 1966]{Hayashi-1966}

Hayashi, C. 1966, ARA\&A, 4, 171
 

\bibitem[Iben 1967]{Iben-1967}

Iben, I. 1967, ARA\&A, 5, 571


\bibitem[Mikheyev \& Smirnov 1986a]{Mikheyev-Smirnov-1986a}

Mikheyev, S. P. \& Smirnov, A. Yu. 1986a, Sov J Nucl Phys, 42, 913


\bibitem[Mikheyev \& Smirnov 1986b]{Mikheyev-Smirnov-1986b}

Mikheyev, S. P. \& Smirnov, A. Yu. 1986b, in Massive Neutrinos in
          Astrophysics and in Particle Physics, Proceedings of the Sixth
          Moriond Workshop, ed. by O. Fackler and J. Tran Thanh Van
          (Gif-sur-Yvette: Edition Fronti$\grave{\rm e}$res), p. 355



\bibitem[Podolak, Hubbard \& Pollack 1993]{Podolak-Hubbard-Pollack-1993}

Podolak, M., Hubbard, W. M. \& Pollack, J. B. 1993, in 
       Protostars and Planets III, ed. by E. H. Levy \& J. I. Lunine
       (Tucson: University of Arizona Press), p. 1109


\bibitem[Sasselov-1996]{Sasselov-1996}

Sasselov, D. D. 1996, astro-ph 9612216


\bibitem[Sekiguchi-Fukugita-1997]{Sekiguchi-Fukugita-1997}

Sekiguchi, M. \& Fukugita, M. 1997,  Institute for Advanced 
         Study preprint IASSNS-AST 97/34



\bibitem[Stothers 1988]{Stothers-1988}

Stothers, R. B. 1988, ApJ, 329, 712



\bibitem[Totsuka 1997]{Totsuka-1997}

Totsuka, Y. 1997, in Proceedings of the Texas Symposium on 
        Relativistic Astrophysics (to be published)


\end{thebibliography}
\end{document}